\documentclass[apj]{emulateapj}
\usepackage{natbib}
\shorttitle{The evolution of And\,XVI}
\shortauthors{Monelli et al.}

\begin{document}

\title{The ISLANDS project I: 
Andromeda XVI, An Extremely Low Mass Galaxy not Quenched by Reionization}

\author{
Matteo, Monelli\altaffilmark{1,2},
Clara E. Mart{\'i}nez-V{\'a}zquez\altaffilmark{1,2},
Edouard J. Bernard\altaffilmark{3},
Carme Gallart\altaffilmark{1,2},
Evan D. Skillman\altaffilmark{4},
Daniel R. Weisz\altaffilmark{5,6},
Andrew E. Dolphin\altaffilmark{7},
Sebastian L. Hidalgo\altaffilmark{1,2},
Andrew A. Cole\altaffilmark{8},
Nicolas F. Martin\altaffilmark{9,10},
Antonio Aparicio\altaffilmark{1,2},
Santi Cassisi\altaffilmark{11},
Michael Boylan-Kolchin\altaffilmark{12},
Lucio Mayer\altaffilmark{13,14},
Alan McConnachie\altaffilmark{15},
Kristen B.~W. McQuinn\altaffilmark{4},
Julio F. Navarro\altaffilmark{16},
}
 
\altaffiltext{*}{Based on observations made with the NASA/ESA Hubble Space Telescope, 
obtained at the Space Telescope Science Institute, which is operated by the 
Association of Universities for Research in Astronomy, Inc., under NASA contract NAS 
5-26555. These observations are associated with program \#13028}

\altaffiltext{1}{Instituto de Astrof\'{i}sica de Canarias, La Laguna, Tenerife, Spain; monelli@iac.es}
\altaffiltext{2}{Departamento de Astrof\'{i}sica, Universidad de La Laguna, Tenerife, Spain}
\altaffiltext{3}{Institute for Astronomy, University of Edinburgh, Royal Observatory, Blackford Hill, Edinburgh EH9 3HJ, UK}
\altaffiltext{4}{Minnesota Institute for Astrophysics, University of Minnesota, 
Minneapolis, MN, USA}
\altaffiltext{5}{Astronomy Department, Box 351580, University of Washington, Seattle, WA, USA}
\altaffiltext{6}{Hubble Fellow}
\altaffiltext{7}{Raytheon; 1151 E. Hermans Rd., Tucson, AZ 85706, USA}
\altaffiltext{8}{School of Physical Sciences, University of Tasmania, Private Bag 37, Hobart 7005, TAS, Australia}
\altaffiltext{9}{Observatoire astronomique de Strasbourg, Universite de Strasbourg, CNRS, UMR 7550, 11 rue de l'Universite, F-67000 Strasbourg, France}
\altaffiltext{10}{Max-Planck-Institut f\"ur Astronomie, K\"onigstuhl 17, D-69117 Heidelberg, Germany}
\altaffiltext{11}{INAF-Osservatorio Astronomico di Teramo, via M. Maggini, 64100 Teramo, Italy }
\altaffiltext{12}{Department of Astronomy, The University of Texas at Austin, 2515 Speedway, Stop C1400, Austin, TX 78712, USA}
\altaffiltext{13}{Institut f\"ur Theoretische Physik, University of Zurich, Z\"urich, Switzerland}
\altaffiltext{14}{Department of Physics, Institut f\"ur Astronomie, ETH Z\"urich, Z\"urich, Switzerland}
\altaffiltext{15}{Herzberg Astronomy and Astrophysics, National Research Council Canada,
5071 West Saanich Road, Victoria, British Columbia, V9E 2E7, Canada}
\altaffiltext{16}{Department of Physics and Astronomy, University of Victoria, PO Box 1700, STN CSC, 
Victoria BC V8W 3P6, Canada}

\date{Received ccx c, cc; accepted cc c, cc}
%
\begin{abstract}

Based on data aquired in 13 orbits of HST time, we present a detailed
evolutionary history of the M31 dSph satellite Andromeda {\sc XVI}, including
its life-time star formation history, the spatial distribution of its stellar
populations, and the properties of its variable stars. And {\sc XVI} is
characterized by prolonged star formation activity from the oldest epochs until
star formation was quenched $\sim$6 Gyr ago, and, notably, only half of the mass
in stars of And {\sc XVI} was in place 10 Gyr ago. And {\sc XVI} appears to be a
low mass galaxy for which the early quenching by either reionization or
starburst feedback seems highly unlikely, and thus, is most likely due to an
environmental effect (e.g., an interaction), possibly connected to a late infall
in the densest regions of the Local Group. Studying the star formation history
as a function of galactocentric radius, we detect a mild gradient in the star
formation history: the star formation activity between 6 and 8 Gyr ago is
significantly stronger in the central regions than in the external regions,
although the quenching age appears to be the same, within 1 Gyr.  We also report
the discovery of 9 RR Lyrae stars, 8 of which belong to And {\sc XVI}. The RR
Lyrae stars allow a new estimate of the distance, $(m-M)_0$= 23.72$\pm$0.09 mag,
which is marginally larger than previous estimates based on the tip of the red
giant branch.  
\end{abstract}



\section{Introduction}\label{sec:intro}

Nearby resolved dwarf galaxies in the Local Group (LG) constitute a compelling
sample to address fundamental open questions about galaxy evolution. The
variety of properties in terms of mass, luminosity, surface brightness, gas
content, chemical evolution \citep[e.g.][]{mcconnachie12}, together with the
possibility to resolve them into individual stars, offer a large number of
observables to investigate how small systems evolved since their formation to
the present time. In particular, the ability to derive quantitative star
formation histories (SFH) based on deep photometry reaching below the oldest main
sequence Turn-Off \citep[TO, ][]{gallart05} allows us to put firm constraints
on the time of the onset and of the end of star formation, which opens the
possibility to constrain the physical mechanisms directly affecting the early
stages of dwarf galaxy evolution. On the one hand, it is expected that both
internal (supernova feedback, e.g., \citealt{maclow99}) and external mechanisms
(e.g., ionizing photons from the first sources; \citealt{ricotti05,susa04})
affect the star formation activity, terminating it at an
early epoch. On the other hand, the environment is also expected to play a
significant r\^ole on small systems orbiting massive primaries (tidal
stirring: \citealt{mayer01a}; ram pressure: \citealt{mayer06}; resonances:
\citealt{donghia09}), which may have a substantial effect in stripping mass
from small systems, again leading to an early cessation of the star formation.

In a series of papers, based on deep HST/ACS photometry within the framework of
the LCID collaboration \citep{monelli10a, monelli10b,hidalgo11,skillman14}, we
have shown that star formation generally continues well past $z \sim 6$ in the
mass regime $M_{\star} \ga 10^6 M_{\odot}$. However, during the last ten years,
our knowledge of the LG has been deeply influenced by photometric surveys that
have brought about unexpected discoveries and new questions. First, the number
of known LG galaxies has more than doubled in a few years only. Starting with
the discovery of the first faint dwarf \citep[also called ``ultra-faint
dwarfs'',][]{willman06} the known satellites of the Milky Way (MW) jumped from
11 (9 bright dSph plus the Magellanic Clouds) to 37 today. These
faint dwarfs extend the spectrum of galactic properties to a regime of very low
mass, low luminosity, and typically low mean metallicity. They are thought to
have formed stars very early on and for a very short period of time
\citep{brown14}, possibly because cosmic reionization might have inhibited
further star formation in this low mass regime.

All currently known Local Group faint dwarfs fit well within this general trend, apart from
one exception. Leo T, discovered as a stellar over-density in the Sloan Data
Release 5, immediately presented a peculiar combination of low mass ($\sim 10^5
M_{\odot}$,  \citealt{ryanweber08}) and young stellar populations ($<$ 200
Myr, \citealt{irwin07}), together with a large fraction of HI gas
\citep{ryanweber08}. Deeper HST data confirmed the extended star formation
activity from the oldest epochs to the present day
(\citealt{clementini12,weisz12}, see also \citealt{dejong08}), which revives
the question of whether cosmic reionization is the actual cause of thee star formation
quenching in the faintest dwarfs. Remarkably, two more galaxies recently
discovered have stellar mass smaller than that of Leo T  but their CMD show
hints of extended star formation until  intermediate epoch: Eridanus II 
\citep{koposov15,  bechtol15} and Hydra II \citep{martin15},  detected in the
Dark Energy Survey and in the Survey for the MAgellanic Stellar History
footprints.

Similarly to what occurred in the MW, the number of known satellites of M31 has
increased considerably in the last few years
\citep{martin09,richardson11,slater11,bell11, martin13a,martin13c}. This was
mainly thanks to the effort of the PAndAS project \citep{mcconnachie09}. The
discovery of And {\sc XVI} was reported in \citet{ibata07}, from MegaCam/CFHT
observations of the M31 surroundings that later would be folded in the PAndAS
survey \citep{mcconnachie09}. And {\sc XVI} is located $\sim 279$ kpc from M31
in the south-east direction. The initial estimate of its luminosity
\citep[$M_V$ = -9.2 mag,][]{ibata07} suggested a relatively bright object. 
However, more recent estimates (Martin et al. 2016, submitted) revised this
value to a significantly fainter value, $M_V$ = --7.6 mag. First estimates based
on the tip of the red giant branch (RGB) indicated a distance
($m-M$)$_0$=23.60$\pm$0.2 mag, corresponding to 525$\pm$50 kpc, though smaller
values have been suggested  \citep[23.39$^{+0.19}_{-0.14}$,][]{conn12}.
Spectroscopic follow-up supports a low mean metallicity, close to [Fe/H] = --2
\citep{letarte09,collins14, collins15}. However, the most distinctive
characteristic of And {\sc XVI}  is its extended SFH, which
continued to $\sim$6 Gyr ago \citep{weisz14a}. The present work is part of 
the ISLAND project (Initial 
Star-formation and Lives of the ANDromeda Satellites), which obtained a total
of 111 HST orbits to study six satellites of M31 (GO 13028, 13739): And {\sc I}, 
And {\sc II}, And {\sc III}, And {\sc XV}, And {\sc XVI}, and And {\sc XXVIII}. In this paper we present a
detailed reanalysis of the data from \citet{weisz14a}, adding information on the
properties of the variable stars population and on the spatial variation of the
stellar populations. In particular, \S \ref{sec:data} presents a brief summary
of  the ACS data used in this work and a detailed presentation of the And {\sc
XVI} CMD. In \S \ref{sec:vars} we present the discovery and analysis of RR
Lyrae (RRL) stars, and we derive a new distance for And {\sc XVI} 
in \S \ref{sec:distance}. \S \ref{sec:sfh} is devoted to
the derivation of the detailed SFH, while \S \ref{sec:radial} presents an 
analysis of the variation of the properties of And {\sc XVI} as a function of
radius, both in terms of SFH and CMD morphology. The discussion of these
results (\S \ref{sec:discussion}) and a summary of the conclusions (\S 
\ref{sec:conclusions}) close the paper.

 \begin{table*}[ht!]
 \begin{center}
 \caption{Log of the observations}
 \begin{tabular}{ccrcc}
 \hline
 \hline
\textit{Image Name} & \textit{Filter} & \textit{Exp. time} & \textit{Date}   & \textit{MJD} \\
                    &                 & \textit{$s$}     & \textit{(UT start)} &  \textit{d-2,400,000} \\
\hline
%
jc1d09upq & $F475W$  & 1,280  &  2013 Nov 20   12:46:13   &  56616.545139  \\
jc1d09urq & $F814W$  &   987  &  2013 Nov 20   13:10:30   &  56616.560301  \\
jc1d09uuq & $F814W$  & 1,100  &  2013 Nov 20   14:13:37   &  56616.604792  \\
jc1d09uyq & $F475W$  & 1,359  &  2013 Nov 20   14:34:55   &  56616.621076  \\
jc1d10wdq & $F475W$  & 1,280  &  2013 Nov 20   23:55:40   &  56617.010037  \\
jc1d10wfq & $F814W$  &   987  &  2013 Nov 21   00:19:57   &  56617.025199  \\
jc1d10xaq & $F814W$  & 1,100  &  2013 Nov 21   01:23:05   &  56617.069701  \\
jc1d10xeq & $F475W$  & 1,359  &  2013 Nov 21   01:44:23   &  56617.085986  \\
jc1d11ywq & $F475W$  & 1,280  &  2013 Nov 21   09:29:27   &  56617.408499  \\
jc1d11yyq & $F814W$  &   987  &  2013 Nov 21   09:53:45   &  56617.423673  \\
jc1d11z1q & $F814W$  & 1,100  &  2013 Nov 21   10:56:55   &  56617.468199  \\
jc1d11z5q & $F475W$  & 1,359  &  2013 Nov 21   11:18:13   &  56617.484483  \\
jc1d12a2q & $F475W$  & 1,280  &  2013 Nov 21   15:52:01   &  56617.674172  \\
jc1d12a5q & $F814W$  &   987  &  2013 Nov 21   16:16:18   &  56617.689334  \\
jc1d12a9q & $F814W$  & 1,100  &  2013 Nov 21   17:23:30   &  56617.736660  \\
jc1d12zzq & $F475W$  & 1,359  &  2013 Nov 21   17:44:48   &  56617.752522  \\
jc1d13b9q & $F475W$  & 1,280  &  2013 Nov 21   23:50:12   &  56618.006245  \\
jc1d13bbq & $F814W$  &   987  &  2013 Nov 22   00:14:29   &  56618.021407  \\
jc1d13caq & $F814W$  & 1,100  &  2013 Nov 22   01:17:39   &  56618.065933  \\
jc1d13ceq & $F475W$  & 1,359  &  2013 Nov 22   01:38:57   &  56618.082218  \\
jc1d14f2q & $F475W$  & 1,280  &  2013 Nov 22   10:59:39   &  56618.471143  \\
jc1d14f4q & $F814W$  &   987  &  2013 Nov 22   11:23:56   &  56618.485872  \\
jc1d14f7q & $F814W$  & 1,100  &  2013 Nov 22   12:27:08   &  56618.530854  \\
jc1d14fbq & $F475W$  & 1,359  &  2013 Nov 22   12:48:26   &  56618.547139  \\
jc1d14feq & $F475W$  & 1,360  &  2013 Nov 22   14:02:46   &  56618.598771  \\
jc1d14fiq & $F814W$  & 1,100  &  2013 Nov 22   14:28:23   &  56618.615056  \\
\hline
 \hline
 \end{tabular}
 \end{center} 
 \label{tab:tab01}
 \end{table*}


\section{Data}\label{sec:data}

\begin{figure}
 \includegraphics[width=9cm]{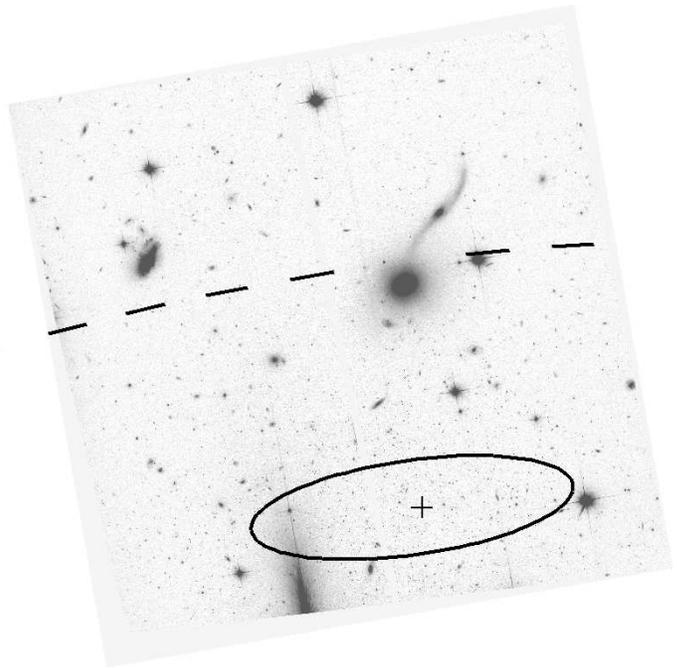}
 \caption{Stacked image of the ACS field on And {\sc XVI} (North is up, East 
 is left). A large number of extended
 sources are clearly visible, and prompted a careful selection of the 
 photometry list. The cross marks the center of {\sc And XVI}, while the solid
 and dashed lines show the ellipses corresponding to r$_e$=1.38r$_h$=1.38$\arcmin$ 
 and r$_e$=5r$_h$=5.00$\arcmin$, being r$_h$ the half-light radius. }
 \label{fig:image}
\end{figure}

\begin{figure}
 \includegraphics[width=9cm]{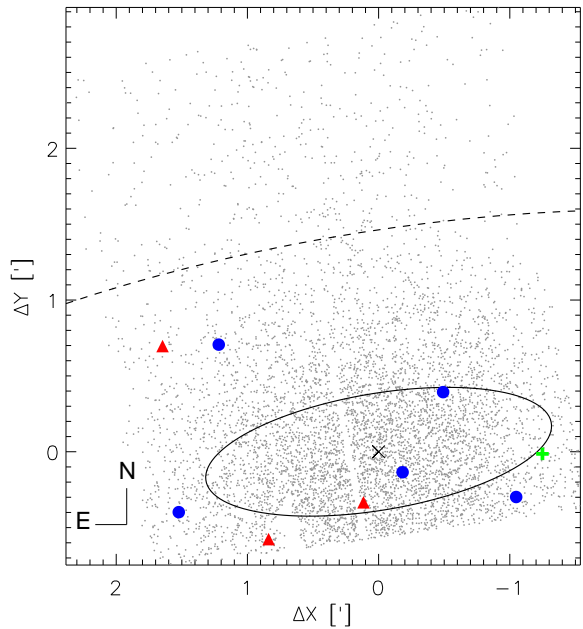}
 \caption{Spatial distribution of bone fide stellar sources in the ACS field around
 And {\sc XVI}. The center of the galaxy is marked by the black cross. The two 
 ellipses  correspond are the same as in Figure \ref{fig:image}. The  global SFH 
 has been derived selecting sources within the dashed line. The location of
 the detected RRL variable stars is shown: red triangles mark the three RR$ab$ 
 type, while blue circles represent the five RR$c$ type stars. The green plus marks
 the position of the peculiar, faint RRL star V0.}
 \label{fig:map}
\end{figure}

The data set used here is the same presented in \citet{weisz14a}, and
consists of 13 ACS images in each the $F475W$ and $F814W$ passbands. Parallel
photometric reductions were conducted using both DOLPHOT and DAOPHOT/ALLFRAME
as was done  for the LCID project \citep[e.g.,][]{monelli10b}.  Here we have
chosen to use the  DAOPHOT/ALLFRAME photometry as a matter of convenience. 
The calibration to the standard VEGAMAG system was done adopting the updated
zero point from the instrument web page. Figure \ref{fig:image} shows a
stacked drizzled image, where a large number of background extended objects
is evident. In particular, note the edge-on galaxy apparently interacting
with the big elliptical to the West, and the group of late-type galaxies in
the North-East. The two ellipses correspond to elliptical radii $r_e$=1.38$\arcmin$,
and 5.00$\arcmin$, and will be used in \S \ref{sec:radial} to investigate the
radial properties. Figure \ref{fig:map} shows the spatial distribution of the
sources in the final catalog.  Big colored symbols mark the position of the 9
discovered RRL stars (see \S \ref{sec:rrl}).

	\subsection{CMD analysis}\label{sec:cmd}

\begin{figure*}[!t]
 \includegraphics[width=17cm,height=15cm]{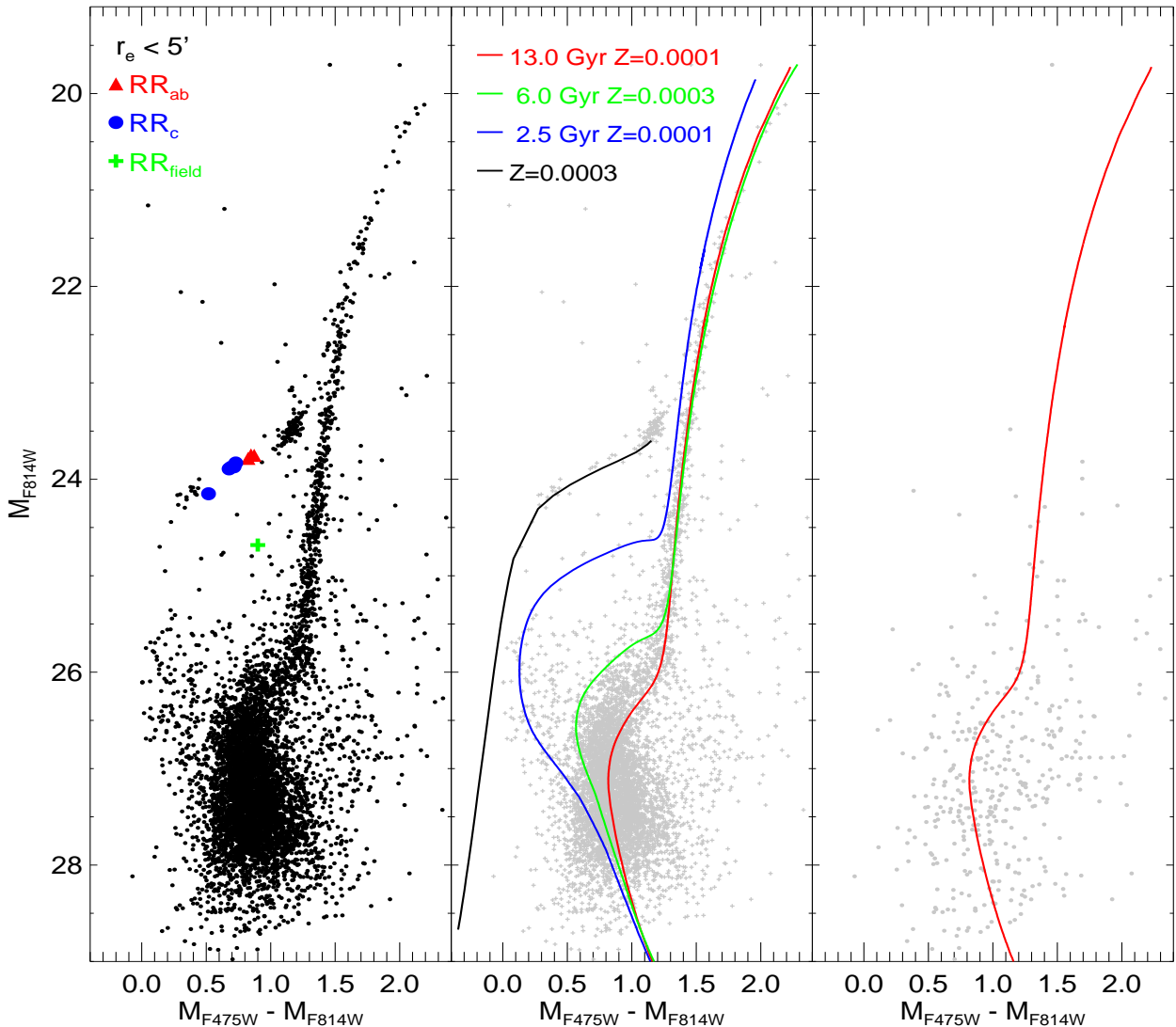}
 \caption{{\em Left -} The CMD of And {\sc XVI}, spanning from the tip of the RGB to well
 below the main sequence TO. Colored symbols show the RRL stars, with the same color code
 as in Figure \ref{fig:map}. {\rm Center - } The same CMD with superimposed selected isochrones
 from the BaSTI database, for labeled age and metallicity. The two selected isochrones
 bracket completely the TO region and the color spread of the RGB, suggesting a significant spread
 both in age and in metallicity. The black line shows the ZAHB for Z=0.0003, which nicely
 reproduce the lower envelope of the HB stars. {\rm Right - }CMD of the outermost
 region in the field of view, for r$_e >$ 5 \arcmin, where the majority of
 the detected sources are polluting unresolved background galaxies. }
 \label{fig:cmd}
\end{figure*}

Figure \ref{fig:cmd} shows the ($F475W - F814W$, $F814W$) CMD of And {\sc
XVI}. In the construction of Figure \ref{fig:cmd} we adopted a reddening of 
E($B$-$V$)=0.06 \citep{schlafly11} and a distance modulus of $(m-M)_0 =$ 23.72 mag. 
The latter value has been derived from the RRL stars, as detailed in 
\S \ref{sec:distance_rrl}. The RRLs discovered in And {\sc XVI} are plotted as 
large symbols and will be discussed in \S \ref{sec:vars}.

A photometric selection was applied according to the sharpness parameter provided 
by DAOPHOT ($|sharp| <$ 0.3). Given the small number of And {\sc XVI} stars,
and the heavy contamination from background galaxies, we performed a further
check on the stacked image, removing a few hundred sources associated with
extended objects and spikes of heavily saturated field stars. Finally, we
ended up retaining 5,714 bona-fide stellar sources within the 5.0\arcmin
 ellipse. These are shown in the CMD of the left panel, where the typical
features of a predominantly old stellar populations clearly appear. The red
giant branch (RGB) spans more than five magnitudes, from the tip at $F814W
\approx$ 20 mag down to $F814W \approx$ 25.5. The horizontal branch (HB) has
a predominantly red morphology  with a well populated red part, concentrated
close to ($F475W - F814W$, $F814W$) $\sim$ (1.2, 23.5) mag, which is well
separated from the RGB, suggesting a limited metallicity spread. On the
other extreme, the HB extends to the blue reaching well beyond the RRL
instability strip to $F475W - F814W$ $\sim$ 0.2 mag. Overall, we
derive an HB morphology index$= -$0.64\footnote{The HB index was introduced
by \citet{lee90} and it is defined as
HBR = ($B$-$R$)/($B$+$V$+$R$), where B and R are the number of HB
stars bluer and redder than the instability  strip, and V is the total
number of RR Lyrae stars.}. 

The central panel of the same figure shows the  comparison with selected
isochrones from the BaSTI\footnote{ http://basti.oa-teramo.inaf.it/index.html}
stellar evolution library \citep{pietrinferni04, pietrinferni09}. In particular,
the red and green lines represent an old (13 Gyr, Z=0.0001) and an
not-too-old-age (6 Gyr, Z=0.0003, \citealt{castellani95}) population. These two
isochrones bracket both the RGB and the main sequence Turn-Off region well.
Interestingly, this suggests that the stellar populations in And {\sc XVI} are
characterized by a  considerable age spread, but a small range of
metallicities. 

Finally, the right panel  presents the sources detected in the outermost region
of the field of view, for r$_e >$ 5\arcmin. The same old isochone as in the
central panel is shown. Roughly 400 sources are present in this diagram, but no
obvious features appear. Many of the detected objects present colors redder than
the MS stars of And {\sc XVI}, suggesting that they are unresolved background
galaxies. Neverthless, we cannot rule out the possibility that some And {\sc
XVI} stars are still present in this region, which will be anyway excluded from
the SFH analysis.

 \begin{table*}[ht!]
 \begin{center}
 \caption{Variable Stars Properties}
 \begin{tabular}{ccccccccccccccc}
 \hline
 \hline
 \textit{Name} & \textit{R.A.} & \textit{Dec.} & \textit{type}  & \textit{P} & \textit{m$_{F475W}$}  & \textit{A$_{F475W}$}    & \textit{m$_{F814W}$}  & \textit{A$_{F814W}$} & \textit{m$_{B}$}  & \textit{A$_{B}$}  & \textit{m$_{V}$}  & \textit{A$_{V}$}   & \textit{m$_{I}$} & \textit{A$_{I}$} \\
\textit{ }     & \textit{hr min sec} & $\arcdeg$ $\prime$ $\arcsec$ & \textit{ }& \textit{d} & \textit{mag}  & \textit{mag} &  \textit{mag}         &  \textit{mag}        & \textit{mag}      & \textit{mag}       & \textit{mag}       & \textit{mag}        & \textit{mag}      & \textit{mag} \\
  \hline
V0 & 00:59:24.38  &  32:22:33.14  &  $ab$  & 0.622 & 25.582 & 0.889 & 24.682 & 0.431 & 25.727 & 0.990 & 25.244 & 0.767 & 24.670 & 0.446 \\
V1 & 00:59:25.33  &  32:22:16.09  &  $c$   & 0.358 & 24.568 & 0.641 & 23.875 & 0.399 & 24.681 & 0.725 & 24.328 & 0.538 & 23.857 & 0.397 \\
V2 & 00:59:27.97  &  32:22:57.56  &  $c$   & 0.391 & 24.560 & 0.557 & 23.831 & 0.284 & 24.668 & 0.616 & 24.308 & 0.454 & 23.812 & 0.290 \\
V3 & 00:59:29.43  &  32:22:25.88  &  $c$   & 0.350 & 24.569 & 0.541 & 23.892 & 0.358 & 24.667 & 0.573 & 24.325 & 0.480 & 23.873 & 0.361 \\
V4 & 00:59:30.84  &  32:22:13.99  &  $ab$  & 0.617 & 24.623 & 0.840 & 23.751 & 0.589 & 24.741 & 0.969 & 24.313 & 0.810 & 23.735 & 0.600 \\
V5 & 00:59:34.27  &  32:21:59.43  &  $ab$  & 0.638 & 24.594 & 1.200 & 23.747 & 0.655 & 24.717 & 1.182 & 24.323 & 0.978 & 23.731 & 0.667 \\
V6 & 00:59:36.07  &  32:23:16.33  &  $c$   & 0.399 & 24.586 & 0.444 & 23.870 & 0.216 & 24.694 & 0.478 & 24.342 & 0.390 & 23.851 & 0.217 \\
V7 & 00:59:37.51  &  32:22:10.07  &  $c$   & 0.288 & 24.668 & 0.300 & 24.150 & 0.161 & 24.736 & 0.316 & 24.495 & 0.251 & 24.134 & 0.157 \\
V8 & 00:59:38.10  &  32:23:15.76  &  $ab$  & 0.651 & 24.608 & 0.673 & 23.783 & 0.427 & 24.734 & 0.719 & 24.327 & 0.574 & 23.767 & 0.432 \\

 \hline
 \hline
 \end{tabular}
 \end{center} 
 \label{tab:tab02}
 \end{table*}


	\subsection{Blue Straggler stars}\label{sec:bss}

\begin{figure}
\vspace{-3cm}
 \includegraphics[width=9cm]{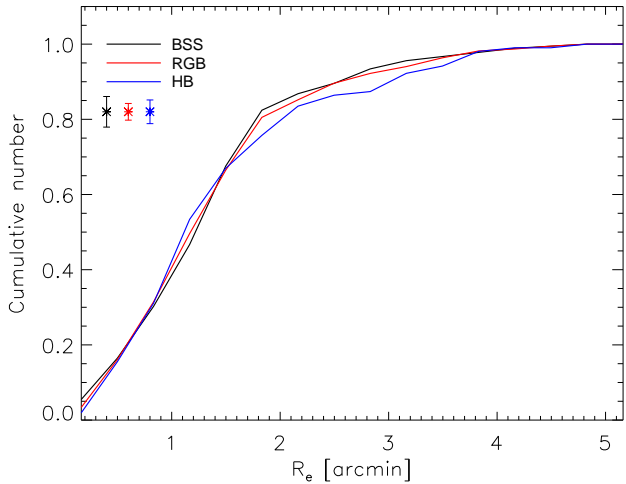}
 \caption{Normalized cumulative radial distribution of stars in the RGB, HB,
 and the candidate BSSs stars. Within the error, no significant differences
 are detected.}
 \label{fig:cumul}
\end{figure}

The CMD clearly show a plume of objects bluer and brighter than the old  MSTO,
between $F814W \sim 25.5$ mag and $F814W \sim 27.5$ mag. They are most likely
Blue Stragglers stars (BSSs) formed by primordial binary stars, as commonly
found in many dSph \citep{mapelli07, mapelli09, monelli12a, santana13}. On the
other hand, stars in that region of the CMD might be genuine young objects, with
ages in the range between $\sim$1 and $\sim$ 3 Gyr. The blue line in Figure
\ref{fig:cmd} represents a metal-poor isochrone of 2.5 Gyr, which provides a
fair agreement with the observed sequence. If And {\sc XVI} hosted such a young
population, one would expect to find it spatially concentrated in the innermost
region of the galaxy, as commonly observed in LG dwarfs. Figure \ref{fig:cumul}
shows the cumulative  distribution of stars in the blue plume, the RGB and the
HB. Within the error, they are identical a a function lf elliptical radius. This
indirectly supports the inference that the stars in the blue plume are BSSs and
not a young population. The plume of blue objects causes a minor peak in
the SFH between 2 and 3 Gyr ago \S \ref{sec:sfh_global},
which contributed with $\sim$3\% of the stellar mass. Both the age range and the
mass percentage are consistent with those estimated in Cetus and Tucana
\citep{monelli12a}. This again indirectly supports the BSS hypothesis.

\begin{figure*}[t]
 \includegraphics[width=17cm]{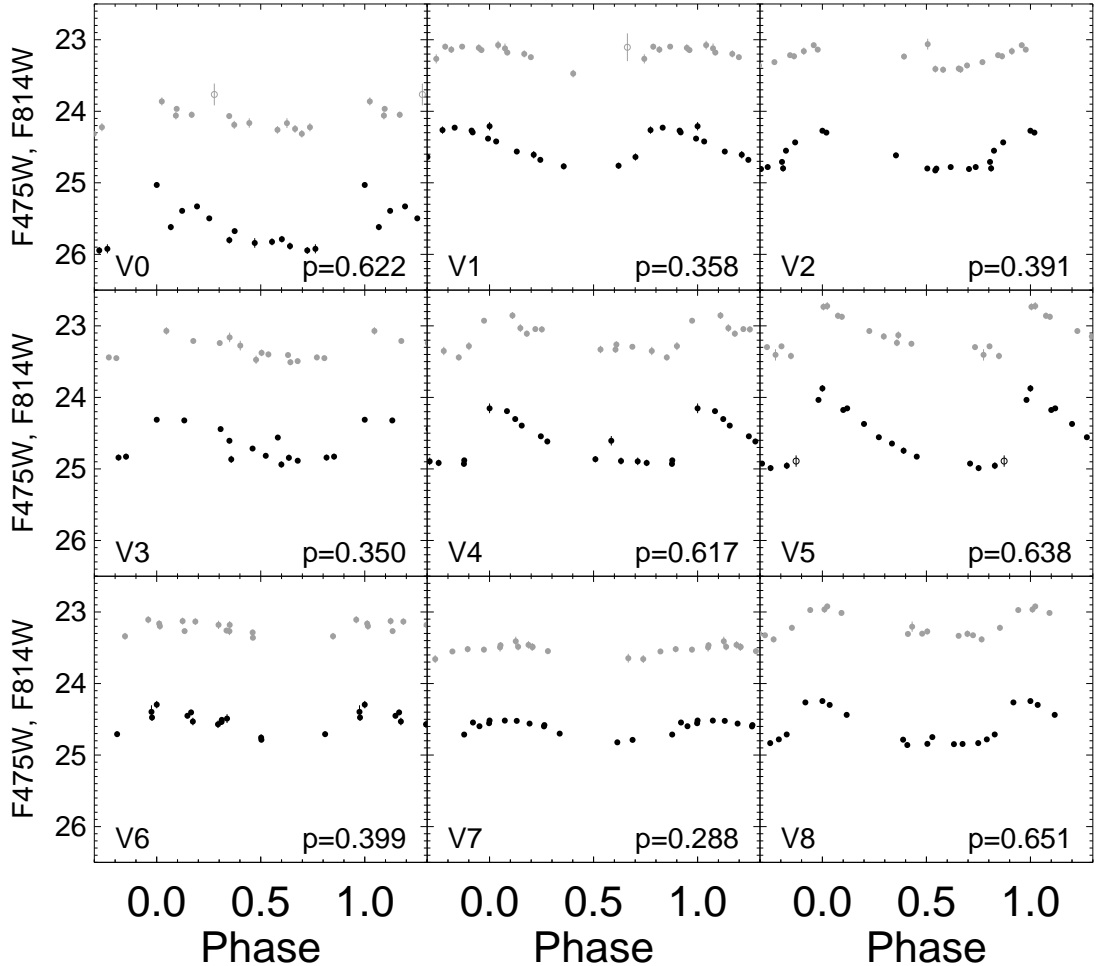}
\vspace{-8cm}

 \caption{The light curves of the 9 discovered RRL stars. In each panel, black
 and grey points refer to the $F475W$ and $F814W$ data. For the sake of 
 clarity, $F814W$ was shifted by -0.6 mag to avoid overlap. Open points are
 excluded from the LC fit due to large photometric error. Note that V0 is
 significantly fainter than the other 8 RRL stars, compatible with an M31 
 field  variable.}
 \label{fig:lcs}
\end{figure*}

\section{Variable stars}\label{sec:vars}

Our observational strategy was designed for optimal time sampling of short 
period ($\lesssim$1 d) variable stars such as
RRL and Anomalous Cepheids (ACs). All the observations were executed within
$\sim$2.1 days, and were organized in six  visits, five of two and one of three
orbits.  Moreover, each orbit was split into one $F475W$ and one $F814W$
exposure,  and a sequence $F475W$-$F814W$-$F814W$-$F475W$ was observed in each
group of two visits. This allowed a larger time difference between the two
images at shorter wavelength, where the amplitude is larger. Table 1 reports
the observation log, listing the image name, filter, exposure time, starting
date of observation and modified Julian date at mid exposure.

	\subsection{RR Lyrae stars}\label{sec:rrl}

Candidate variable stars were identified following the same approach adopted for
the galaxies of the LCID project \citep{bernard09,bernard10,bernard13}. In
particular, we used the variability index introduced by \citet{stetson96b}. The
light curves of selected candidates were individually visually inspected, and
nine variables were confirmed.  Given their pulsational properties and the
location on the CMD, we classify all of them as RRL stars.  The $F475W$ and
$F814W$ magnitudes were re-calibrated to the Johnson $BVI$ system using the same
relations adopted in \citet{bernard09}. Table 2 summarizes the properties of the
confirmed variables which are named in order of increasing right ascension.
Their position in the CMD is shown in Figure \ref{fig:cmd}: red triangles and
blue circles represent  RR$ab$ and RR$c$ type stars, respectively.
Interestingly, one variable (V0) is significantly fainter (green plus), by
$\sim$0.8 mag in the $F814W$ band. Figure \ref{fig:lcs} presents the light
curves of the nine variables. Despite the small number of phase points, the time
sampling chosen when preparing the observations provides a fairly homogeneous
coverage of the light curves. In particular, we do not find any obvious problem
with V0 that may account for the fainter magnitude, though the light curve is
admittedly noisy. We have checked whether a significantly higher metal content
may be responsible for such a lower luminosity. Adopting the
luminosity-metallicity relation by \citet{clementini03}, we derive that an
[Fe/H] approximately solar is required to explain such a large magnitude
difference. Such a large metallicity spread within the population of stars able
to form RRL in And {\sc XVI} looks unlikely, in particular if we take into
account the relatively small range estimated  both spectroscopically
\citep{letarte09, tollerud12} and with the SFH (see \S \ref{sec:sfh_global}).
Alternatively, we can assume that V0 does not belong to And {\sc XVI}, and the
magnitude difference is due to a distance effect. Assuming a metal content
of [Fe/H]=$-$1.9 and the metallicity-luminosity relation from
\citet[][see \$ \ref{sec:distance_rrl}]{clementini03}, we derive a distance
difference between V0 and the rest of variables of the order of $\sim$290 kpc.
Given that And {\sc XVI} is located $\sim$200 kpc closer than M31, this means
that V0 is compatible with being located  $\approx$ 100 Kpc beyond M31, but
still well within its virial radius, thus being a possible candidate M31 halo
star \citep{ibata14a}.

\begin{figure}
 \includegraphics[width=9cm]{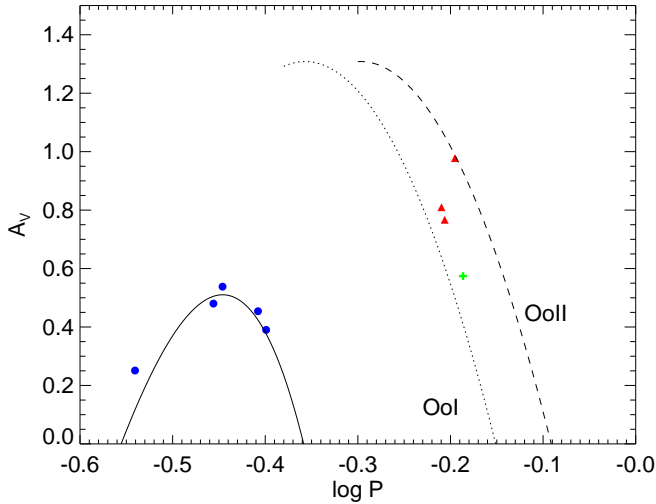}
 \caption{Period-Amplitude (Bailey) diagram for the nine detected variables. The dotted
 and dashed lines are the loci of RRL stars in Oosterhoff I and Oosterhoff II type
 globular clusters, from \citet{cacciari05}. The solid line is the analogous curve for RR$c$
 type from \citep{kunder13c}. }
 \label{fig:bailey}
\end{figure}

Figure \ref{fig:bailey} shows the period-amplitude (Bailey) diagram for the
detected variable stars. The dotted and dashed lines mark the loci of Oosterhoff
I and Oosterhoff II globular clusters, respectively, from \citet{cacciari05}. 
The solid line is the locus defined for RR$c$ stars, from \citet{kunder13c}. The
three And {\sc XVI} RR$ab$ stars occupy the region intermediate to the two
curves. However, both  the mean period of RR$ab$ stars ($<P_{ab}>$=0.636 d) and
of the RR$c$  type ($<P_c>$=0.357 d) are close to the typical values for the 
Oosterhoff II type stellar systems.  Note that the ratio between the number of
RR$c$ and RR$ab$ is unusually large \citep{catelan09}, and And {\sc XVI} is the
only dwarf known with more RR$c$ than RR$ab$. This finding is particularly
intriguing given the red morphology of the HB which would favor the sampling of
the red part of the instability strip, where the RR$ab$ are located. However it
might  be possibly related to the small total number  of RRL stars.
Alternatively, this  effect could be related to the low metallicity of the
oldest stars, such as the RR Lyrae stars which would be preferentially located
in the blue part of the HB. However, we note that other M31  satellites with
similar small number of RRL variables  and bluer HB morphology such as And XI
and And XIII \citep{yang12}, the  number of RR$ab$ type is larger than that of
RR$c$ type (10 vs 5 and 8 vs 1, respectively).

	\subsection{Anomalous Cepheids}\label{sec:ac}
	
We report that we did not discover any AC in the surveyed
area of And {\sc XVI}. This kind of pulsating variables, present only in
metal-poor \citep[Z$<$0.0006][]{fiorentino06} populations, are centrally
He-burning stars typically $\sim$1 mag brighter then RRL stars. They can form
through two different channels: {\em i)} single, evolved stars of mass 1.3
$\lesssim$ M $\lesssim$ 2.2 M$_{\odot}$, therefore younger than $\sim$1 Gyr;
{\em ii)} coalescent binary stars evolved after the BSS phase.  Despite 
the fact that ACs
have been observed in many dSph galaxies (Sculptor: \citealt{kaluzny95};
Fornax: \citealt{bersier02}; Carina: \citealt{dallora03,coppola13}, Draco:
\citealt{kinemuchi08}; Cetus and Tucana: \citealt{bernard09}), the
non-detection in And {\sc XVI} is not surprising, and we ascribe it to its low
mass. First, the lack of ACs agrees with the lack of recent star formation, 
thus excluding the first formation channel.
Second, the small number of stars populating the blue plume of BSS ($\sim230$)
implies that very few evolved stars of this population are expected. These,
due to their mass, tend to occupy the red part of the HB, at temperatures
lower than the instability strip, and few such stars are clearly visible above
the red HB, at $F814W\sim$23 mag. Finally, adopting the relation between the
frequency of ACs and the luminosity of the host galaxy, discovered by
\citet{mateo95} and updated by \citet{fiorentino12b}, we estimate that
1$\pm$1 ACs are expected in the surveyed area, in agreement with the 
current observations.

\section{Distance estimate}\label{sec:distance}

		\subsection{RR Lyrae distance estimate}\label{sec:distance_rrl}

Pulsational properties of RRL stars can be used to derive a robust estimate
of the distance. In the following we will use three different methods. In the
analysis, we did not include the faint V0 RRL star.

{\em a)} First, we adopt the relation between the intrinsic luminosity, $M_V$, and
the metallicity. In the range below [Fe/H] = --1.6 we assume two linear
relations\footnote{the $F475W$ and $F814W$ magnitudes were re-calibrated to the
Johnson $BVI$ system using the same relations adopted in \citet{bernard09}}:

\begin{equation}
M_V(RR) = 0.866 (\pm0.085) + 0.214(\pm0.047)\hbox{[Fe/H]}
\end{equation}

from \citet{clementini03} and

\begin{equation}
M_V(RR) = 0.72 (\pm0.07) + 0.18(\pm0.07)\hbox{\rm [Fe/H]}
\end{equation}

from \citet{bono03}. We assume a value for the [Fe/H]=$-$2, in agreement with
the  available spectroscopic measurements \citep{letarte09,collins14, collins15}.  For
the metal content, the \citet{clementini03} and the \citet{bono03} relations
provide absolute magnitude values of $M_V$ = 0.438 and 0.360 mag, respectively.
We derive absolute distance moduli for And {\sc XVI}, corrected for
extinction, of $(m-M)_0$ = 23.72$\pm$0.09  mag and 23.79$\pm$0.08 mag,
respectively, corresponding to 554 and 572 kpc. We note that a change in the
metal content by 0.2 dex affects the distance estimates by $\sim$0.04 mag

{\em b)} It is well established that RRL stars obey a period-luminosity-metallicity 
relation in the near-infrared, which can be expressed  in the form
\begin{equation}
Mag = a + b[Fe/H] +cLogP
\end{equation}

We adopt here the most updated theoretical relations from Marconi et al. (2015), 
both for the Wesenheit W($I$,$B-I$) and W($I$,$V-I$) magnitudes. 

We used here the full sample of RRL stars after fundamentalizing the
RR$c$ type by adding 0.127 to the
logarithm of their period. We calculated the Wesenheit apparent magnitudes
of each star, and adopting these relations we derived the true distance modulus.
Assuming [Fe/H] $\sim$ -2.3 dex (Z $\sim$ 0.0001) the two relations provide
$(m-M)_0$ = 23.74$\pm$0.03 mag and $(m-M)_0$ = 23.77$\pm$0.06 mag,
respectively. A slightly larger metallicity, [Fe/H] $\sim$ -1.8 dex (Z
$\sim$ 0.0003) shortens the derived distance by few hundredths of magnitude:
$(m-M)_0$ = 23.68$\pm$0.03 mag and $(m-M)_0$ = 23.70$\pm$0.03
 mag.

{\em c)} An independent method to derive the distance based on the RRL
properties was introduced by \citet{caputo00b} and takes advantage of the
period-luminosity-metallicity relation at the first overtone blue edge of the
instability strip. Ideally, this method works well if the blue side of the
instability strip is  well sampled, which is not the case for the current data
set. However, the few RR$c$ type stars found can provide an upper limit to
the distance. Applying the relations from \citet{caputo00b} to the 
shortest period star, we derive a distance modulus of 23.83$\pm0.07$, 
again assuming [Fe/H]=$-$2.

Overall, these different methods applied to the RRL stars sample of And {\sc
XVI} provide consistent results about its distance. For consistency with
previous analysis of isolated galaxies within the framework of the LCID
project, we will adopt  the distance derived with the $M_V$-[Fe/H] relation by
\citet{clementini03}, $(m-M)_0$= 23.72$\pm$0.09 mag,  to derive the SFH in
\S \ref{sec:sfh}.

\begin{figure}
 \includegraphics[width=9cm]{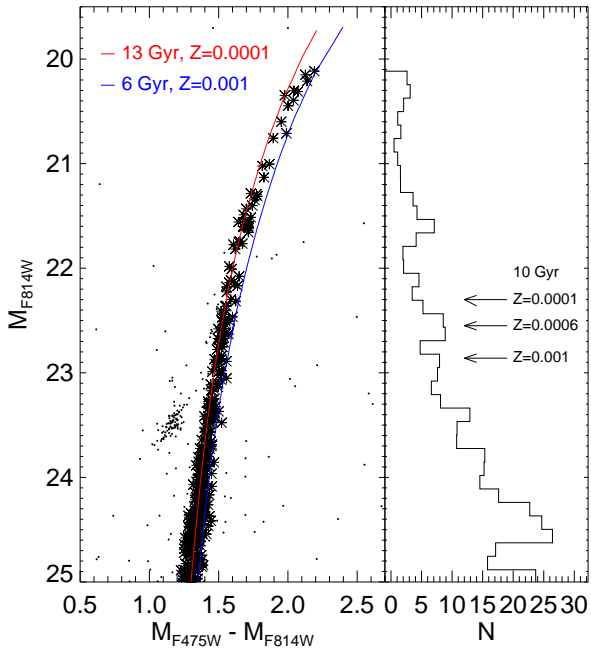}
  \caption{{\em Left - }Colour-Magnitude diagram of And {\sc XVI} showing the
 RGB stars (asterisks) used to detect the RGB tip. Two isochrones are also
 over-plotted: Z=0.0001, t=13 Gyr (red line), Z=0.001, t=6yr (blue line). We
 assumed $(m-M)_0 = 23.72$ mag and E(B-V) = 0.06 mag. {\rm Right -
 }Luminosity function of RGB stars. The three arrows mark the expected position
 of the RGB bump for three isochrones of 10 Gyr: Z=0.0001, Z=0.0006, and Z=0.001.}
 \label{fig:tip}
\end{figure}

	\subsection{RGB tip distance estimate}\label{sec:distance_rgb}

\citet{ibata07} estimated the distance of And {\sc XVI} to be (m-M)$_0$ =
23.6$\pm$0.2 mag (525 Kpc), based on the position of the tip of the RGB. 
A more recent study by \citet{conn12}, based on a more sophisticated analysis 
of the same feature,
suggested a slightly shorter distance, (m-M)$_0$ = 23.39$^{+0.19}_{-0.14}$ (476
kpc). We note that the distance estimate based on the RRL stars are systematically
larger than those based on the tip of the RGB. However, they are is still in 
agreement, within the error bars, with the value provided by \citet{ibata07}, 
and only in marginal agreement at the 2$\sigma$ level with the measurement
by \citet{conn12} .

Figure \ref{fig:tip} summarizes our attempt to derive a distance to And {\sc
XVI} based on tip of the RGB as detected in the ACS data. The left panel shows
a zoom of the CMD in the RGB region, and the right one presents the luminosity
function of RGB stars in the $F814W$  band. These are highlighted  by big
asterisks in the left panel. The plot clearly shows that the region of the RGB
tip is heavily under-sampled, with only 9 stars detected in the half brightest
magnitude. This is far from the at least 50 stars recommended by
\citet{madore95} to derive a distance modulus with 0.1 uncertainty. This is
also supported by the comparison with theoretical isochrones (red line:
Z=0.0001, t=13 Gyr; blue: Z=0.001, t=6 Gyr). Assuming (m-M)$_0$=23.72 mag (from
the RRL estimate, see \S \ref{sec:distance_rrl}), it is evident that the
brightest portion of the RGB is devoid of observed stars. Note that a shorter distance
modulus would move the isochrones to brighter apparent magnitudes, thus
worsening the problem. Given the little contamination from both And {\sc XVI}
AGB and foreground field stars, we can set an upper limit to the distance,
assuming that the brightest observed star is representative of the tip. This
has magnitude $F814W$ = 20.116 mag. The F814W absolute magnitude of the RGB tip
shows a mild dependence on  the metallicity in the metal regime appropriate for
the stars in  And {\sc XVI}. In more detail, theoretical predictions based on
BaSTI stellar  models show that $M_{F814W}^{tip}$ is equal to -4.087 at
Z=0.0001 and to  -4.166 for Z=0.001. When combining these model predictions with
an  extinction estimate of $A_{F814W}=0.11$ mag, we obtain a distance  modulus
upper limit ranging from 24.09 to 24.17 mag, i.e., in the range  657 - 682 Kpc. A
visual inspection of the CMD from \citet{ibata07} discloses at least one very
bright star is missing in our photometry, possibly because it is outside our
field of view. This is probably what causes the difference  in the derived
distance using the same approach. In any case, it is evident that the poor
statistics in the RGB star  counts are strongly hampering the possibility to use
the RGB tip method  for a robust distance estimate.

In passing, we note that for the same reason no clear detection of the RGB bump
is possible. The luminosity function in the right panel of Figure \ref{fig:tip}
does not show any clear evidence of the RGB bump. The three overplotted arrows
mark the position of the RGB bump derived from theoretical isochrones of 10 Gyr,
and metallicity ranging from Z=0.0001 to Z=0.001. We note that an observed peak
around $F814W\sim$22.5 mag agrees well with the predicted bump for Z=0.0006 and 
age of 10 Gyr. As the bump positions depends both on the age (fainter bump
for increasing age) and the metallicity (fainter bump for increasing metallicity)
there is some degree of degeneracy. However, it seems clear that the observed peak cannot
be reproduced with very metal-poor populations, as the predicted bump
for Z=0.0001 and an age of 10 Gyr is too bright ($F814W\sim$22.35 mag), 
and gets brighter for decreasing age, while it gets virtually undetectable
for older ages, as the magnitude extension of the loop drops. Similarly, in the case of
more metal-rich populations, the predicted bump is too faint, and an age of 6 Gyr is 
required fit the observed peak (though with a color that is too red).

\section{Star formation history}\label{sec:sfh}

\begin{figure}
 \includegraphics[width=9cm]{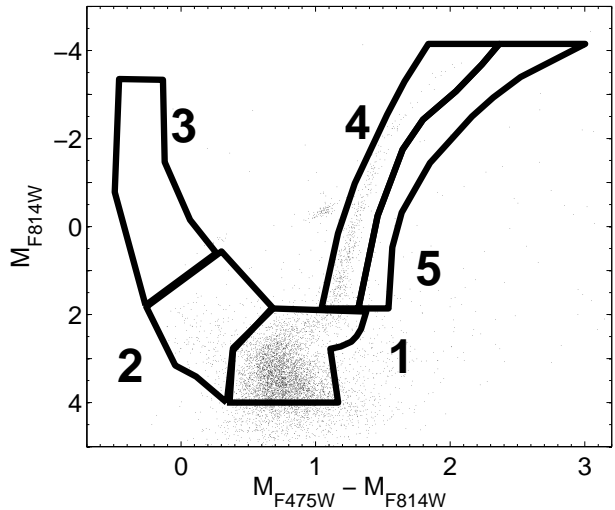}
 \caption{CMD of And {\sc XVI} with superimposed the five regions ({\it bundles}) used to
 derive the SFH. See text for details.}
 \label{fig:bundles}
\end{figure}

\begin{figure*}
 \includegraphics[width=17cm]{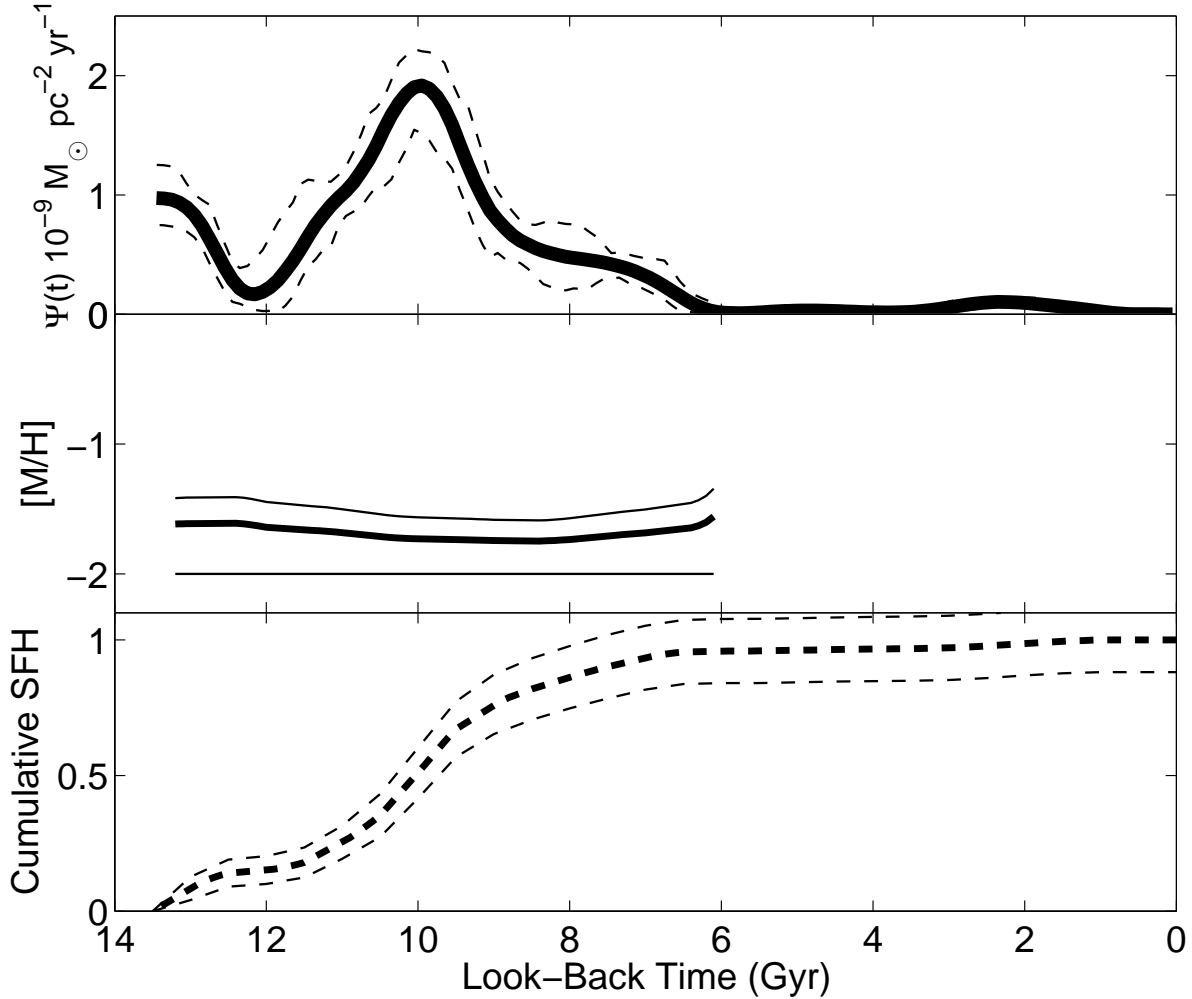}
 \caption{SFH of And {\sc XVI}. As a function of look-back time, from top to
 bottom the three panels show the star formation rate, the age-metallicity relation,
 and the cumulative SFH. Clearly, And {\sc XVI} was able to sustain star formation
 for at least 6 Gyr.}
 \label{fig:sfh}
\end{figure*}

	\subsection{Star formation history derivation}\label{sec:sfh_parameters}
	
The SFH was derived using the IAC-star, MinnIAC and IAC-pop codes
\citep{iacstar,hidalgo11,iacpop}, in a similar fashion as already presented in previous
papers of the LCID project \citep{monelli10a,
monelli10b,hidalgo11,skillman14}. For the present data set, we used a model
CMD of 3$\times$10$^6$ stars with ages and metallicities uniformly distributed
in the ranges 0 $<$ t $<$ 13.5 Gyr and 0.0001 $<$ Z $<$ 0.0025. Observational
errors were simulated taking into account the results of 2$\times$10$^6$ artificial
stars.

IAC-star requires the selection of a number of parameters that are used in the
solution derivation. On the one hand, parameters used to build the model CMD such
as the amount of binary stars and the initial mass function were chosen to be the
same as in previous LCID papers. Namely, we used a 40\% binary faction ($q>$0.5) and 
the \citet{kroupa02} IMF (x=1.3 for $M<M_{\odot}$ and  x=2.3 for $0.5 < M_{\odot} <
100$). To run IAC-pop and MinnIAC, decisions have to be taken concerning the 
parametrization of both the age and metallicity bins (that define the 
``simple stellar populations'') and that of the CMDs. 
The adopted age and metallicity bins were: \\ 
age=[0 1 2.5:1:13.5] Gyr \\ 
metallicity=[0.0001 0.0003 0.0005 0.0007 0.0010 0.0015 0.0020 0.0025]\\ 
The sampling of the CMD is based on macro-regions, called {\it
bundles} (see Figure \ref{fig:bundles}). In each bundle, stars are counted in a 
regular grid of boxes, whose size is fixed and constant.
The main limit of the current data set is the relatively small number of stars in 
the observed CMD, which  can introduce noise in the solution if too fine of a sampling 
of the CMD is adopted. Therefore, we performed a number of tests to optimize the 
bundle and boxe sizes. We found that the final solution is mostly affected by two
factors: {\em i)}  the sizes of the boxes in {\it bundle 1}; {\em ii)} the
inclusion of the RGB in {\it bundle 4}. 

Most of the information on the age of the stellar population comes from the main
sequence and TO region. For predominantly old populations as those present in
And {\sc XVI}, most of the information will come from {\it bundle 1}. {\it
Bundle 3} and {\it bundle 5} are useful to set limits to the youngest
populations and the highest  metallicity, respectively, while {\it bundle 2}
samples the blue plume. 
In previous works, in order to give more weight
to the TO region, we adopt smaller boxes in the corresponding bundle. However,
we found that, in comparison with our previous LCID experience, we had to
significantly increase the size of individual boxes in this bundle in order to
avoid fluctuations and the appearance of spurious populations in the solution.
Namely, the box size chosen is (color, magnitude) = (0.04, 0.2) mag, compared
to typically (0.02, 0.1) in LCID. Given the little or negligible number of
stars, larger boxes are used in bundles 2, 3 and 5. The HB is excluded from
the SFH analysis because the details of its morphology depend on highly
unknown factors such as the mass loss during the RGB phase, and they are not 
properly modeled in our synthetic CMD.

The second major difference with the LCID strategy is that including the RGB
significantly improves the solution as well. With the LCID galaxies we had
demonstrated that, whenever the CMDs are well populated by at least tens of
thousands of stars, the inclusion of the RGB has little, if any, effect on the
final solution, and typically the $\chi^2$ increases \citep{bernard12}. This is
mostly due to the fact that the age is highly degenerate in the RGB, while a
bundle such as the current {\it bundle 5} is always useful to set a constraint
to the most metal-rich population. In the current analysis, where only few
thousand stars are available, we found that the solution strongly benefits from
the inclusion of a bundle on the RGB. The main effect is that spurious
populations (such as simultaneously very old and very metal-rich ones) disappear
from the solution.

\begin{figure}
 \includegraphics[width=9cm]{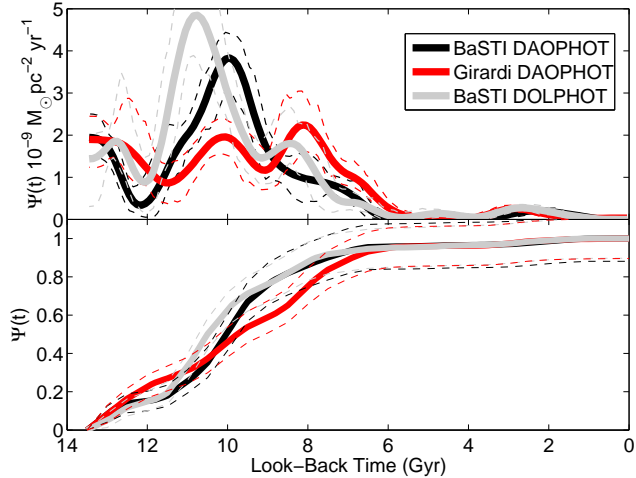}
 \caption{SFH solutions obtained adopting different photometry sets and 
 different stellar evolution libraries. The top panel shows the SFR as a function
 of time, while the bottom one present the normalized cumulative SFH.
 }
 \label{fig:sfr_comparison}
\end{figure}

	\subsection{Global star formation history of And {\sc XVI}}\label{sec:sfh_global}
	
The SFH of And {\sc XVI} was derived using only stars within 5 $r_h$ from
the center. The total number of stars used to derive the SFH are 3985, 202, and
491, in {\it bundles 1,2, 4} respectively. For larger galactocentric distances,
the majority of sources are expected to be background unresolved galaxies, 
Nevertheless, the comparison with theoretical isochrones in Figure \ref{fig:cmd}
suggests that a small fraction of And {\sc XVI} stars may be present at larger
radius. Scaling the number of objects in the outer regions found in the same
bundles within 5$r_h$, we find that an upper limit of $\sim$4.5\% of
contaminating objects may be affecting the star counts, thus not strongly
affecting the derived SFH. In particular, since the distribution of the contaminating
galaxies in the CMD does not resemble that of a stellar population, we do not
expect that they originate any strong features at a specific age in the SFH.

The final solution is presented in Figure \ref{fig:sfh}. The three panels
represent, from top to bottom, the star formation rate (SFR), the
age-metallicity relation (AMR) and the cumulative SFH as a function of the
look-back time. And {\sc XVI} is populated by both old and intermediate age
stars. It started forming stars at the oldest possible epoch. Remarkably, in our
SFH solutions, there appears to be a significant very old peak at 13.5 Gy ago,
followed by a sudden drop of star formation. After a minimum occurred 12 Gyr
ago, star formation increased again and reached its peak $\sim$10 Gyr ago. This
is an extremely interesting finding, as this feature is not common either among
the MW dSph satellites, nor the isolated ones such as Cetus and Tucana. In fact,
they typically present one single dominant event of star formation occurred at
the oldest epochs \citep[see e.g.,][]{monelli10b, monelli10c, deboer12a,
deboer12b}. The second distinctive feature we recover, as already  found by
\citet{weisz14a}, is that the star formation activity extends for many Gyr,
vanishing 6 Gyr ago.  The blue plume of stars in {\it bundle 2} produces the
small peak at $\sim$3 Gyr, which we interpret as BSS stars (see \$
\ref{sec:bss}). We also recover a fundamentally  constant AMR, with metallicity
not exceeding [M/H]=$-$1.5 (Z=0.0006), in agreement with the qualitative
comparison with theoretical isochrones.  
further constrain the nature 
cumulative SFH reveals that 10\% of And {\sc XVI}  stellar mass was in place by
$z\sim$6 ($\sim$12.8 Gyr ago), that is when the reionization epoch concluded,
and that And {\sc XVI} formed 50\% of its stellar mass by  $z\approx$2, or
$\sim$10.1$\pm$0.2 Gyr ago (see Table 3 for the derived integrated and mean
quantities).

 \begin{table}[ht!]
 \begin{center}
 \caption{Integrated and mean quantities}
 \begin{tabular}{lc}
 \hline
 \hline
\textit{Quantity} & \textit{value} \\
                    &              \\
\hline
%
$\int<\Psi(t)>dt$ (10$^6 M_{\odot}$) & 1.92$\pm$0.03 \\
$<\Psi(t)>$ (10$^{-8} M_{\odot} yr^{-1} pc^{-1}$) & 3.6$\pm$0.1 \\
$<$age$>$ (Gyr) & 9.9$\pm$0.1 \\
$<$[Fe/H]]$>$ 10$^{-4}$ dex & 4.2$\pm$0.1 \\
 &  \\
\hline
 \hline
 \end{tabular}
 \end{center} 
 \label{tab:tab03}
 \end{table}


Figure \ref{fig:sfr_comparison} presents a comparison between the SFH recovered
using different photometry sets and stellar evolution libraries. In particular,
together with the previous DAOPHOT+BaSTI solution (black lines), we show the
SFH obtained with DAOPHOT+Girardi (red lines, \citealt{girardi00}) and DOLPHOT+BaSTI (grey lines),
The Figure presents both the SFR as a function of time (top panel) as well as
the normalized cumulative SFH (bottom). The plots disclose a general
very good agreement. In particular, the three solutions confirm the fundamental
results that the star formation in And {\sc XVI} did extend to $\sim$ 6 Gyr
ago, and that there is no dominant initial event as in other dSph such as Cetus
and Tucana. We exclude that this can be an artifact due to photometric errors, as
they are too small to affect the TO morphology causing the age spread, in either
photometry. Moreover, the three solutions confirm an initial star
formation followed by a less intense activity. In particular, the use of either
photometry set together with the BaSTI models provide a minimum at 12 Gyr, while
the subsequent maximum is 1 Gyr younger in the DOLPHOT+BaSTI solution than in
the DAOPHOT+BaSTI one. Interestingly, while the BaSTI solution provides a 
strong peak at such ages, the solution based on the Girardi library is
characterized by a flatter SFR, though the age of the peaks agrees very well with
the BaSTI solutions. The consistency between the three solutions is clear in the
bottom panels, where the cumulative SFHs agree at the 1-$\sigma$ 
level.

\begin{figure}
 \includegraphics[width=9cm]{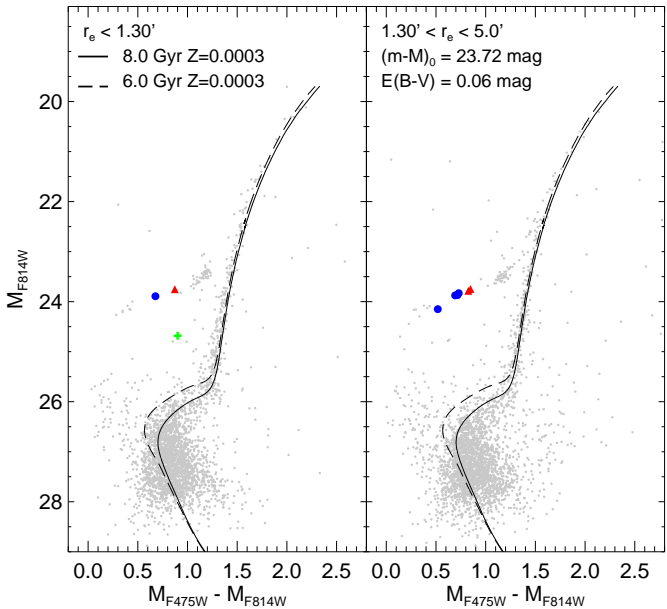}
  \caption{CMDs of the inner (r$_e < $1.38\arcmin, left) and outer regions
 (1.3$\ge$ r$_e <$5.00\arcmin, right) of And {\sc XVI}. Colors
 symbols show the RRL stars in each region. The separation between the inner
 and outer region is such that  the two CMDs contain the same number of
 sources within the {\it bundles}. Two not-too-old isochrones are
 over-plotted (Z=0.0003, t=6,8 Gyr). The number of stars in the TO region
 comprised by the two curves is large in the inner than in the outer region,
 suggesting stronger star formation at this age in the  closer to the center
 of the galaxy.
 }
 \label{fig:grad_cmd}
\end{figure}

\begin{figure}
 \includegraphics[width=9cm]{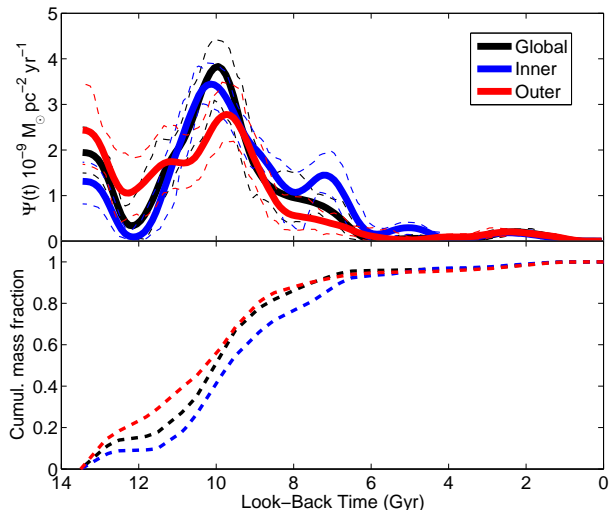}
 \caption{The SFR derived for the inner and outer region of And {\sc XVI}.
 The star formation was slightly more prolonged in the inner than in the outer region,
 but no strong gradient was found.}
 \label{fig:grad_sfh}
\end{figure}

\section{Radial spatial gradient}\label{sec:radial}

In this section we investigate how the properties of And {\sc XVI} change as
a function of the distance from its center. First, we note that we do not
have a symmetric spatial sampling of the galaxy. In fact, due to a bright
field star next to the innermost regions of And {\sc XVI}, we were forced to
point the telescope such that the center of the galaxy is next to the edge of
the ACS camera, at (X,Y)$\approx$(566,1847) px (see the black cross in Figure
\ref{fig:map}). Second, we estimate that the current ACS data cover
$\approx$23\% of the galaxy area.

For the following analysis, we take advantage of a homogeneous derivation of
the structural parameters of all M31 dwarf spheroidal galaxies that fall in
the PAndAS footprint \citep{salomon15} and use the following,
updated values for the centroid (0:59:30.3+-0.4;+32:22:34+-0.4), ellipticity
(0.29$\pm$0.08), position angle (98$\pm$9\arcdeg), and half-density radius
(1.0$\pm$0.1\arcmin).We calculated the elliptical distance for each star from
the galaxy center, and we used it to select  three regions.  The two panels
of Fig. \ref{fig:grad_cmd} show the CMD of the inner and  outer regions, selected such that they have a
similar total number of sources {\it in the bundles used for the SFH
derivation} ($\sim$2,300). This occurs at r=1.38$r_h$. Interestingly, the
overall morphology of the CMD does not change strongly as a function of
radius. In the following we analyze in detail the differences in the SFH,
and  how these reflect in the variation of the CMD morphology. The CDM of the
outer regions, already presented in the right panel of Figure \ref{fig:cmd},
clearly demonstrates that there is marginal evidence for the presence of And
{\sc XVI} stars beyond 5r$_h$ (r$_e$=5.0\arcmin).

		\subsection{The spatially and temporally extended SFH of And {\sc XVI}}\label{sec:sfh_rad}

To guide the eye, we over-plotted on Figure \ref{fig:grad_cmd} two isochrones
from the BaSTI database, assuming Z=0.0003 and ages= 6,8 Gyr. Comparing the
two panels we found that the region between two curves is
slightly more populated  in the inner (276 stars) than in the outer region
(191 stars),  suggesting that the star formation rate $\sim$6 Gyr ago was
higher in the inner than in the outer region. It also may indicate that the
star formation was slightly more prolonged toward  the
center of And {\sc XVI}, as commonly found in  nearby dwarf galaxies, though
the effect looks small. It is remarkable that And {\sc XVI} was able to
sustain star formation for at least 6 Gyr over a vast fraction of its main
body. 

To support this finding, we derive the SFH in the two elliptical regions, in
identical way as for the full galaxy. The results are shown in Figure
\ref{fig:grad_sfh}, where the calculated SFRs vs time are over-plotted. The
figure shows that the main features are consistent in the inner, outer, and the
global solutions. The SFR in both the central and external region presents an
initial peak followed by a decreased activity. The main peak is recovered at
similar age ($\sim$10 Gyr ago), and star formation continues to 6 Gyr in both
regions. However, at the most recent epochs, it presents stronger activity in
the central  part compared to the outskirts, with a secondary peak occurred
$\sim$7 Gyr ago. It must be stressed that the uncertainties are large, mostly
due to the small number of stars used to derive both solutions,  and therefore
such detailed comparison should be treated cautiously. However, the fact that
And {\sc XVI} was able to sustain star formation for at least 6 Gyr over its
entire body remains a solid result. This is significantly different from what
was found in other dwarfs. For example, the spatial variation of the SFH in LGS
3 and Phoenix \citep{hidalgo13} indicate the presence of a gradient in the age
of the youngest populations, which are confined in the central regions only.
Similar conclusions have been reached also in the case of the MW satellites
Fornax and  Carina \citep{deboer13,deboer14b}, which are dominated by
intermediate-age  populations in the center and by purely old populations in
the outskirts.

\section{Discussion}\label{sec:discussion}

\begin{figure*}
 \includegraphics[]{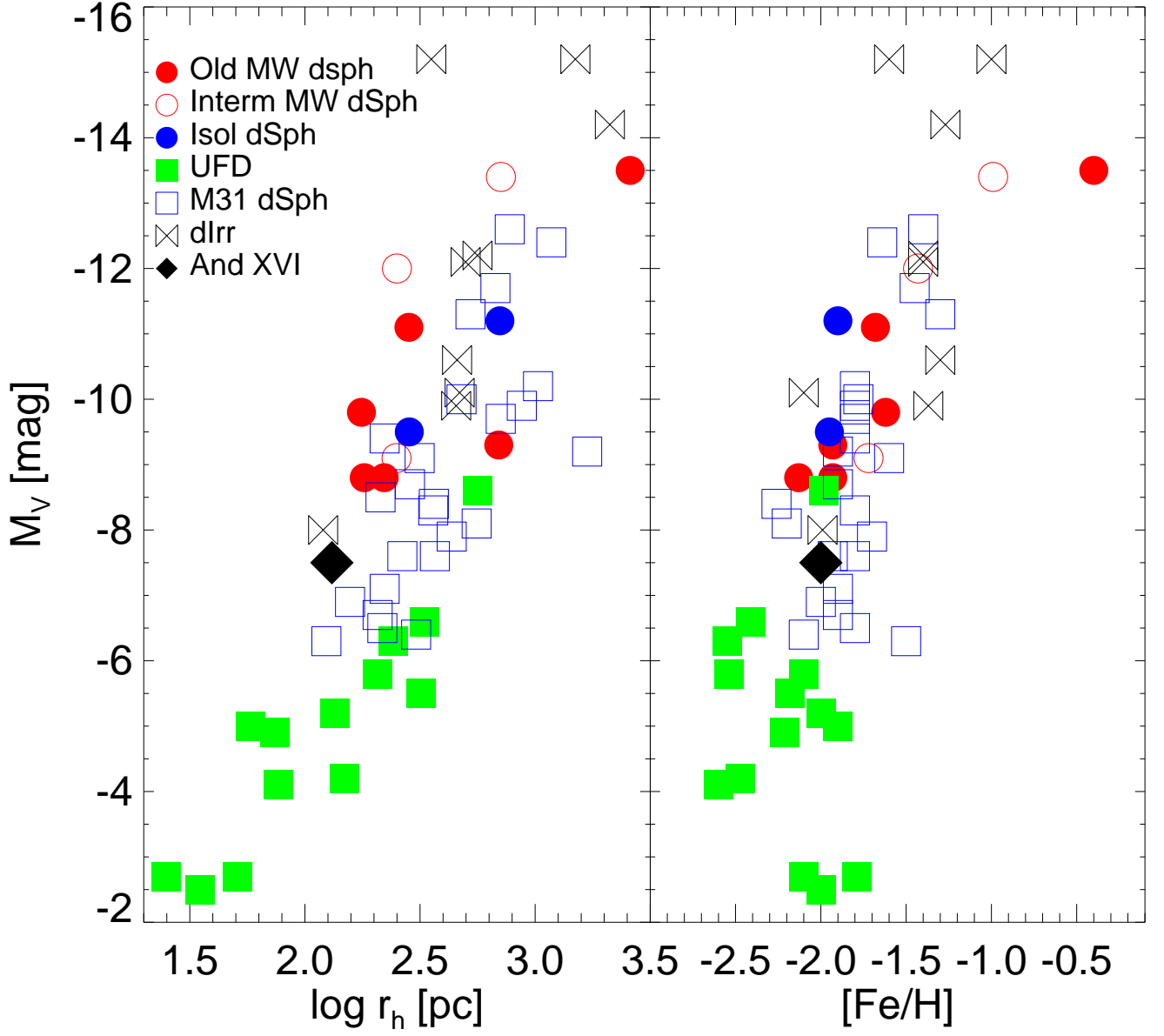}
 \caption{M$_V$ magnitude vs the logarithm of the half-light radius (left) and
 the metallicity (right) for  LG dwarf galaxies of different morphological type.
 The data are from \citet{mcconnachie12}, but with updated values for And {\sc XVI}.
 Different symbols indicate galaxies of
different morphological types: red circle: MW dSphs (full: purely old systems;
open: systems with strong intermediate populations); blue circles: isolated dSph
(Cetus and Tucana); green squares: MW faint dwarfs satellites; open squares: M31 dSph
satellites; butterflies: dIrr systems (including transition types such as LGS3
and Phoenix).}
 \label{fig:relations}
\end{figure*}

Given its size and luminosity, And {\sc XVI} is somewhat at the boundary between
classical and faint dwarfs. Figure \ref{fig:relations} shows the absolute $M_V$
magnitude of Local Group dwarfs as a function of their size (half light radius,
$r_h$) and metallicity. The data are from the compilation paper by
\citet{mcconnachie12}, and the plots partially replicate his figures 6 and 12
(see also \citealt{clementini12}, their figure 1). Different symbols indicate
LG dwarf galaxies of different morphological types, as labeled. We updated here
the position of And {\sc XVI}, shown as a black  diamond, using the luminosity
from Martin et al 2016 (submitted). And {\sc XVI}  occupies the faint tail of the
M31 satellites sequence, being $\sim$1 mag brighter than M31 dwarfs of similar
size, such as And XI and And XX. With respect to previous estimates
\citep{ibata07}, the absolute $M_V$ magnitude increased by $\sim$1.7 mag,
moving And {\sc XVI} significantly closer to the faint dwarfs region ($M_V$ =
$-$7.5 mag), but nonetheless it is still $\sim $2-3 mag brighter than Galactic
faint dwarfs of similar size such as Leo V and Ursa Major II.

And {\sc XVI} is thus a small mass satellite of M31, located relatively far
from both its host ($\sim$279 kpc) and the MW ($\sim$575 kpc). The most
striking feature of its evolution is that it was able to sustain star formation
for $\sim$7 Gyr and, as proven in the previous section, over most of its body,
with only a small spatial gradient in the sense that the youngest star formation
(6-8 Gyr ago) was stronger in the inner regions. This occurrence is an
interesting and peculiar feature among LG dwarfs. In fact, broadly speaking, it
is something intermediate between the two typical observed behaviors. 
Following the nomenclature introduced by \citet{gallart15}, we identify that 
the majority of dSph galaxies are {\itshape fast} systems, i.e., they have formed 
stars for a short amount of time at the oldest epochs (e.g., Draco, Ursa Minor, 
Cetus, Tucana). On the other extreme, {\itshape slow} dwarf galaxies which present
current or recent star formation are characterized by continuous activity from
the oldest to youngest epochs (e.g., Leo A: \citealt{cole07}; Leo T: 
\citealt{weisz12,clementini12}; DDO210: \citealt{cole14}; the Fornax dSph: 
\citealt{deboer12b,delpino13}; the Magellanic Clouds: \citealt{smeckerhane02,noel09,
meschin14}). Within this scheme, the dominant old peak of star formation makes
And {\sc XVI} similar to a {\itshape fast} system, but nonetheless the extended activity
is typical of {\itshape slow} galaxies, though the quenching occurred $\sim$ 6 Gyr ago.
What mechanisms influenced the evolution of And {\sc XVI}? What favored the extended
star formation, and what caused its termination?

We derived that the mass formed in the
surveyed area during the first two Gyr is of the order of $\approx 3\times10^4
M_{\odot}$ (15\% of the total mass). Therefore, And {\sc XVI} would have
properties comparable to a typical faint dwarf, if star formation had been truncated
at a similar epoch. This suggests that, despite the similar stellar mass back then,
And {\sc XVI} was not strongly affected by reionitazion, which is thought to be
the strongest mechanism shutting down star formation in low mass Milky Way satellites
\citep{brown14}. On the other hand, the properties of the old
population in And {\sc XVI} are reminiscent of those of the old population in
the low-mass dIrr isolated galaxies, Leo A and Leo T, at least in terms of
integrated quantities. On the one hand, the mean star formation rate of Leo A
between 13.5 and 11.5 Gyr ago was $\sim$2.$\times10^{-5} M_{\odot} yr^{-1}$,
implying that this dIrr formed in the first 2 Gyr a mass of stars of the order of
4$\times$10$^4$ M$_{\odot}$. This is within a factor of 2 of what was produced
by And {\sc XVI}\footnote{Taking into account the area covered by ACS data and
the size of Leo A\citep{vansevicius04}, we estimate that this number might be
underestimated by a factor or $\approx$2-3, thus not affecting the following
discussion}. Moreover, the number of RRL stars is very similar in both systems
(8 $v_s$ 10, \citealt{bernard13}). On the other hand, Figure
\ref{fig:relations} also shows that, in both planes, And {\sc XVI}  is located
remarkably close to Leo T, the lowest mass star forming galaxy known in the LG.
In particular, And {\sc XVI} is $\sim $0.5 mag fainter than Leo T which,
despite its low mass (total mass $< 10^7 M_{\odot}$, \citealt{simon07}, 
$M_{\star} \sim 1.2\times10^5 M_{\odot}$ \citealt{ryanweber08}, thus comparable
to that of And {\sc XVI}), was able to form stars over a Hubble time
\citep{weisz12,clementini12}. 

This suggests that the initial properties of And {\sc XVI}, Leo A, and Leo T
were similar to those of a faint dwarf progenitor. Nonetheless, if the initial
masses were similar, And {\sc XVI}, Leo T, and Leo A would have been equally
vulnerable as faint dwarfs to the quenching effect of reionization. This clearly
does not seem to be the case, since in the SFH there is no trace of a strong
damping effect during the early evolution, contrary to what occurs in  faint
dwarfs.  {\itshape Possibly, this is indicating that the different evolution is
dictated by the environmental conditions}. At present, Leo T is located in
relative isolation quite similar to  And {\sc XVI}, at $\sim$400 kpc from the MW
and more than $\sim$900 kpc from  M31. Interestingly, the negative radial
velocities of both Leo T and And {\sc XVI} with respect of both spirals and the
LG barycenter is compatible with them approaching the LG for the first time.
Leo A is remarkably one of the most isolated systems at the
fringes of the LG. Together with DDO210 and VV124 it belongs to the restricted
group of dwarf galaxies that did not ever strongly interact with either the MW
and M31 all along their history \citep{mcconnachie12}. {\itshape The similarity
of And {\sc XVI} and these dIrrs may  also indirectly support the idea that And
{\sc XVI} was initially located in a lower density environment, far from both
the ionizing radiation and the gravitational effect of the growing MW and M31,
thus explaining the prolonged  star formation despite the initial low mass. This
has been proposed to be  generally the case for {\itshape slow} systems
\citep{gallart15}. }

Moreover, it has been suggested that And {\sc XVI} is among the least dark
matter dominated of the M31 satellites \citep{collins14}. This as well might  be
an indication of a slower mass assembling history maybe related to the formation
in a low-density environment. Although such small systems are expected to be
strongly affected by reionization, the subsequent evolution may be driven by a
complex interplay of mass assembly history, effect of the reionization and
effect of stellar feedback. Theoretical models by \citet{benitezllambay14}
suggest that the stellar feedback acts as regulator of the evolution of small
galaxies after the reionization epoch: in those systems where the star formation
started before the reionization, the stellar feedback contributes to sweep out
the gas, causing a definitive termination in the star formation. In those
systems where no stars formed before the reionization epoch, this contributes to
heat up and disperse the gas, but is not strong enough to permanently remove the
gas from these systems. This gas is later recollected by the central halo and can
start producing stars mostly at intermediate to young ages.  Leo A, Leo T and
And {\sc XVI} may fit in this scheme, and therefore they may be galaxies with
mass below threshold for star formation before the reionization.

\section{Conclusions}\label{sec:conclusions}

We have presented a detailed analysis of the And {\sc XVI} dSph galaxy, 
satellite of M31, based on deep CMD obtained from ACS data. The main
conclusions can be summarized as follows: \\

$\bullet$ We have derived three SFH of And {\sc XVI} using two different
photometric reduction (DAOPHOT and DOLPHOT) and two stellar evolution 
libraries (BaSTI and Girardi), obtaining a very good agreement independently
on the assumptions; \\
$\bullet$ The SFH of And {\sc XVI} at the oldest epochs seems different from both the MW and
isolated dSph, as the dominant peak occurred relatively late, around 10 Gyr ago,
is preceded by an initial peak at the oldest
ages, followed by a period of decreased activity; \\
$\bullet$ Despite the low stellar mass (M$\sim$10$^5$M$_{\odot}$), And {\sc XVI}
presents an extended star formation activity, which begun at the oldest 
epochs and was maintained until $\sim$6 Gyr ago; \\
$\bullet$ We detected 9 variable stars, all RRL stars. Eight of them
belong to And {\sc XVI}, while one is compatible with being a more distant,
M31 halo field star; \\
$\bullet$ We provided a new estimate of the distance of And {\sc XVI},
$(m-M)_0$= 23.72$\pm$0.09 mag, based on the properties of RRL stars. We 
found that different methods (Luminosity-metallicity relation, 
period-luminosity-metallicity relation) provide values slightly larger than 
previous estimates based on the RGB tip; \\
$\bullet$ We discussed the properties of And {\sc XVI} in comparison
with other LG dwarfs. And {\sc XVI} occupies the faint end of the dSph
sequence. However, we found that if its star formation would have been 
truncated 12 Gyr ago, today it would closely resemble a faint dwarf galaxy
in stellar mass. \\
$\bullet$ The SFH of And {\sc XVI} is consistent with a formation and early
evolution in a low-density environment, which favored a slow mass assembly
and prolonged star formation. A late arrival in the inner region of the LG
may have been the cause of the termination in star formation occurring $\sim$7 Gyr ago.

New data available for more M31 satellites, collected within the framework 
of this project, will allow us to build a fundamental sample to compare
the MW, M31, and isolated dwarfs in the LG.

\section*{acknowledgments} 

We are grateful to the anonymous referee for the pertinent comments which improved
the paper. The authors thanks M. Marconi and V. Braga for providing the coefficients of
the period-luminosity relations. MM is grateful to G. Fiorentino and to G.
Bono for the discussion on the HB morphology and the RRL properties. Support
for this work has been provided by the Education and Science Ministry of
Spain (grants AYA2013-42781, AYA2014-56765-P). DRW is supported by NASA through 
Hubble Fellowship grant HST-HF-51331.01 awarded by the Space Telescope Science 
Institute. MBK is supported by the HST grants AR-12836 and AR-13888.
%
%

{\it Facility:} \facility{HST (ACS)}


\end{document}